\documentclass{emulateapj}
\usepackage{apjfonts}
\usepackage{color}

\newcommand{\pivec}{\mbox{\boldmath $\pi$}}
\newcommand{\muvec}{\mbox{\boldmath $\mu$}}
\newcommand{\te}{t_{\rm E}}
\newcommand{\thetae}{\theta_{\rm E}}
\newcommand{\pie}{\pi_{\rm E}}
\newcommand{\pien}{\pi_{{\rm E},N}}
\newcommand{\piee}{\pi_{{\rm E},E}}

\definecolor{darkbrown}{RGB}{139,69,19}

\shorttitle{OGLE-2016-BLG-1469}
\shortauthors{HAN ET AL.}
\begin{document}

\title{ OGLE-2016-BLG-1469L: Microlensing Binary Composed of Brown Dwarfs}

\author{
C.~Han$^{1}$, A.~Udalski$^{2,22}$, T. Sumi$^{3,23}$, A.~Gould$^{4,5,24}$,
\\
and\\
M.~D.~Albrow$^{6}$, S.-J.~Chung$^{4,7}$, Y.~K.~Jung$^{8}$,   Y.-H.~Ryu$^{4}$, I.-G.~Shin$^{8}$,   
J.~C.,~Yee$^{8}$,   W.~Zhu$^{5}$,      S.-M.~Cha$^{4,9}$, S.-L.~Kim$^{4,7}$, D.-J.~Kim$^{4}$,    
C.-U.~Lee$^{4,7}$,    Y.~Lee$^{4,9}$,   B.-G.~Park$^{4,7}$   \\ 
(The KMTNet Collaboration),\\
I.~Soszy{\'n}ski$^{2}$, P.~Mr{\'o}z$^{2}$, P.~Pietrukowicz$^{2}$, M.~K.~Szyma{\'n}ski$^{2}$, 
J.~Skowron$^{2}$, R.~Poleski$^{2,5}$, S.~Koz{\l}owski$^{2}$, K.~Ulaczyk$^{2}$, M.~Pawlak$^{2}$\\
(The OGLE Collaboration),\\
F.~Abe$^{10}$, Y. Asakura$^{10}$, D.~P.~Bennett$^{11,12}$, I.~A.~Bond$^{13}$, A.~Bhattacharya$^{12}$, 
M.~Donachie$^{14}$, M.~Freeman$^{14}$, A.~Fukui$^{15}$, Y.~Hirao$^{16}$, Y.~Itow$^{10}$,
N.~Koshimoto$^{16}$, M.~C.~A.~Li$^{14}$, C.~H.~Ling$^{13}$, K.~Masuda$^{10}$, Y.~Matsubara$^{10}$,
Y.~Muraki$^{10}$, M.~Nagakane$^{16}$, K.~Ohnishi$^{17}$, H.~Oyokawa$^{10}$, 
N.~J.~Rattenbury$^{14}$, To.~Saito$^{18}$, A.~Sharan$^{14}$, D.~J.~Sullivan$^{19}$, D.~Suzuki$^{11,12}$, 
P.~J.~Tristram$^{20}$, T.~Yamada$^{3}$, T.~Yamada$^{20}$, A.~Yonehara$^{12}$, R.~Barry$^{21}$\\
(The MOA Collaboration)\\
}
\affil{$^{1}$   Department of Physics, Chungbuk National University, Cheongju 28644, Republic of Korea}
\affil{$^{2}$   Warsaw University Observatory, Al. Ujazdowskie 4, 00-478 Warszawa, Poland}
\affil{$^{3}$   Department of Earth and Space Science, Graduate School of Science, 
                Osaka University, Toyonaka, Osaka 560-0043, Japan}
\affil{$^{4}$   Korea Astronomy and Space Science Institute, Daejon 34055, Republic of Korea}
\affil{$^{5}$   Department of Astronomy, Ohio State University, 140 W. 18th Ave., Columbus, OH 43210, USA}
\affil{$^{6}$   University of Canterbury, Department of Physics and Astronomy, Private Bag 4800, 
                Christchurch 8020, New Zealand}
\affil{$^{7}$   Korea University of Science and Technology, 217 Gajeong-ro, Yuseong-gu, Daejeon 34113, Republic of Korea}
\affil{$^{8}$   Harvard-Smithsonian Center for Astrophysics, 60 Garden St., Cambridge, MA, 02138, USA}
\affil{$^{9}$   School of Space Research, Kyung Hee University, Yongin 17104, Republic of Korea}
\affil{$^{10}$  Institute for Space-Earth Environmental Research, Nagoya University, Nagoya 464-8601, Japan}
\affil{$^{11}$  Code 667, NASA Goddard Space Flight Center, Greenbelt, MD 20771, USA}
\affil{$^{12}$  Deptartment of Physics, University of Notre Dame, 225 Nieuwland Science Hall, Notre Dame,
                IN 46556, USA}
\affil{$^{13}$  Institute of Natural and Mathematical Sciences, Massey University, Auckland 0745, New Zealand}
\affil{$^{14}$  Department of Physics, University of Auckland, Private Bag 92019, Auckland, New Zealand}
\affil{$^{15}$  Okayama Astrophysical Observatory, National Astronomical Observatory of Japan, 3037-5
                Honjo, Kamogata, Asakuchi, Okayama 719-0232, Japan}
\affil{$^{16}$  Department of Earth and Space Science, Graduate School of Science, Osaka University,
                Toyonaka, Osaka 560-0043, Japan}
\affil{$^{17}$  Nagano National College of Technology, Nagano 381-8550, Japan}
\affil{$^{18}$ Tokyo Metropolitan College of Aeronautics, Tokyo 116-8523, Japan}
\affil{$^{19}$ School of Chemical and Physical Sciences, Victoria University, Wellington, New Zealand}
\affil{$^{20}$ Mt. John University Observatory, P.O. Box 56, Lake Tekapo 8770, New Zealand}
\affil{$^{21}$ Astrophysics Science Division, NASA Goddard Space Flight Center, Greenbelt, MD 20771, USA}
\footnotetext[22]{The OGLE Collaboration.}
\footnotetext[23]{The KMTNet Collaboration.}
\footnotetext[24]{The MOA Collaboration.}

\begin{abstract}
We report the discovery of a binary composed of two brown dwarfs, based on the 
analysis of the microlensing event OGLE-2016-BLG-1469.  Thanks to detection of 
both finite-source and microlens-parallax effects, we are able to measure both 
the masses $M_1\sim 0.05\ M_\odot$, $M_2\sim 0.01\ M_\odot$, and distance 
$D_{\rm L} \sim 4.5$ kpc, as well as the projected separation $a_\perp \sim 0.33$ 
au.  This is the third brown-dwarf binary detected using the microlensing method, 
demonstrating the usefulness of microlensing in detecting field brown-dwarf binaries 
with separations less than 1 au.
\end{abstract}


\section{Introduction}
 
Studies about brown dwarfs are important because the masses of brown dwarfs occupy the gap between 
the least massive star and the most massive planets and thus they can provide important clues in 
understanding the formation mechanisms of both stars and planets \citep{Basri2000}. In addition, brown 
dwarfs may be as abundant as stars in the Galaxy.

Considering that multiplicity of stars is a ubiquitous result of star formation process, an 
important fraction of brown dwarfs may reside in binaries. Since the first discoveries by 
\citet{Rebolo1995} and \citet{Nakajima1995}, there have been numerous discoveries of 
brown dwarfs. See the archives of brown-dwarf candidates maintained by 
C.~Gelino et al.\ (http://DwarfArchives.org) 
and J.~Gagne.\ (https://jgagneastro.wordpress.com/list-of-ultracool-dwarfs/).  
However, the number of known binary systems composed of brown dwarfs, 
e.g.\ \citet{Luhman2013} and \citet{Burgasser2013}, is small.\footnote{There exist 52 
brown-dwarf binaries among 1281 brown dwarfs listed in Gelino Catalog.}
The studies on the binary properties 
such as the binary frequency and the distributions of separations and mass ratios between 
binary components are important to probe the nature of brown dwarfs because these properties 
are influenced by both the process of the formation and dynamical interaction between the components 
within systems. Therefore, a sample composed of an increased number of binary brown-dwarf systems 
will be important to give shape to the details of the brown-dwarf formation process.

Microlensing can provide a useful tool to search for binary brown-dwarf systems, especially 
those that are difficult to detect using other methods. Due to their extremely low luminosities, 
detecting brown dwarfs using either the imaging or spectroscopic methods is restricted to 
those in nearby stellar associations within which brown dwarfs are young enough to emit 
observable light. By contrast, microlensing, which is caused by the gravity of an intervening 
object (lens) between an observer and a background star (source), occurs regardless of the 
lens brightness, and thus it can be used to detect old field brown-dwarf populations. 
Furthermore, while the spectroscopic measurement of a brown-dwarf mass is extremely 
difficult \citep{Zapatero2004}, the chance to measure the mass of a binary lens is 
relatively high, enabling to make the brown-dwarf nature of the object definitive. 
The usefulness of the microlensing method has been demonstrated by the detections and the 
mass measurements of two field brown-dwarf binaries \citep{Choi2013}.

In this paper, we present the microlensing discovery of another binary system composed of 
brown dwarfs. Coverage of the caustic-approach region and the long timescale of the lensing 
event enable to uniquely determine the mass of the lens system and thus to identify the 
substellar nature of the binary components.

The paper is organized as follows. In Section~2, we describe observations of the lensing 
event and the data acquired from them. In Section~3, we explain the detailed procedure of 
modeling the observed lensing light curve and present the solution of the lensing parameters. 
In Section~4, we present the physical parameters of the lens system. 
In Section~5, we discuss the nature of the low relative lens-source proper motion 
estimated from the light curve analysis. 
We summarize the results of the analysis and conclude in Section~6.

\begin{figure*}[t]
\epsscale{0.8}
\plotone{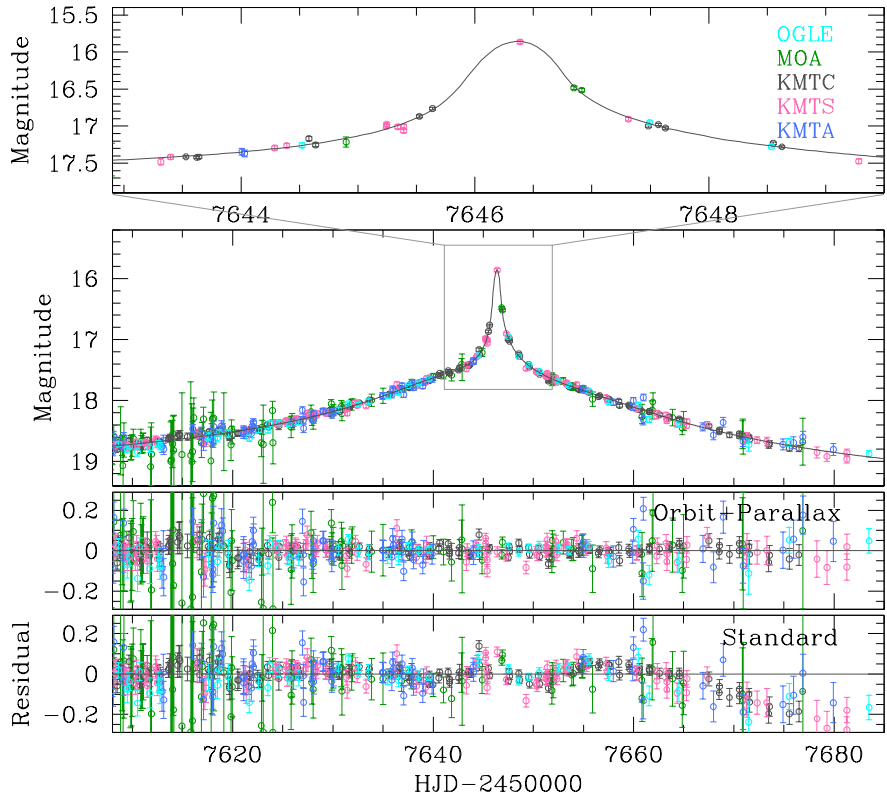}
\caption{
Light curve of the microlensing event OGLE-2016-BLG-1469. The upper panel shows the 
enlarged view of the anomaly around the peak. The two lower panels 
show the residual from the binary-lens models with (orbit+parallax) and without 
(standard) considering higher-order effects.
}
\label{fig:one}
\end{figure*}

\section{Observation and Data}

The brown-dwarf binary was discovered from the observation and analysis of the microlensing 
event OGLE-2016-BLG-1469. The source star of the event is located toward the Galactic bulge 
field with equatorial coordinates 
$(\alpha,\delta)_{\rm J2000}=(18^{\rm h}07^{\rm m}46^{\rm s}\hskip-2pt.98,
-26^{\circ} 17' 23.6'')$, which 
correspond to the Galactic coordinates $(l,b)=(4.75^\circ,-2.92^\circ)$.

The lensing-induced brightening of the source star was first noticed on 2016 July 31 
(${\rm HJD}'={\rm HJD}-2450000\sim 7600.5$) by the Early Warning System \citep[EWS:][]{Udalski2003} of the 
Optical Gravitational Lensing Experiment \citep[OGLE:][]{Udalski2015}. The OGLE lensing 
survey is conducted using the 1.3m telescope located at the Las Campanas Observatory 
in Chile. The OGLE telescope is equipped with a $1.4\ {\rm deg}^2$ field of view (FOV) camera. 
Most images of the OGLE data were taken in the standard Cousins $I$ band with roughly 5\% 
observations in the Johnson $V$ band for color measurement.

The event was also observed by the Microlensing Observations in Astrophysics 
\citep[MOA;][]{Bond2001, Sumi2003}. The MOA survey uses the 1.8m telescope 
at the Mt.~John University Observatory in New Zealand. The MOA telescope is equipped 
with a $2.2\ {\rm deg}^2$ FOV CCD camera. The MOA survey uses a customized wide $R$-band filter 
where the wavelength range corresponds to the sum of the standard Cousins $R$ and $I$ bands. 
In the online list of MOA transient events, the lensing event was dubbed MOA-2016-BLG-542.

The event was also in the footprint of the Korea Microlensing Telescope Network 
\citep[KMTNet:][]{Kim2016} survey. The survey uses globally distributed three identical 
1.6m telescopes located at the Cerro Tololo Inter-American Observatory in Chile (KMTC), 
the South African Astronomical Observatory in South Africa (KMTS), and the Siding Spring 
Observatory in Australia (KMTA).  A $4.0\ {\rm deg}^2$ FOV camera is mounted on each of the 
KMTNet telescope. Most of the KMTNet data were obtained in the standard Cousins $I$-band 
filter with occasional $V$-band observations.

Figure~\ref{fig:one} shows the light curve of the lensing event OGLE-2016-BLG-1469 
with the data taken by 5 telescopes of the 3 lensing surveys. Since the first 
discovery of the event by the OGLE group, the light curve continued to rise until 
a lensing magnification reached $A\sim 20$ at the peak (${\rm HJD}\sim 7646$). 
Covering the peak region of a high-magnification event is important because the 
efficiency to detect planetary signals is high \citep{Griest1998}. An anomalous 
signal actually occurred on ${\rm HJD}'\sim 7646.5$
and the MOA group issued an anomaly alert to the microlensing community for possible 
follow-up observations. In the upper panel of Figure~1, we present the zoom of the 
anomaly. Responding to the alert, 8 images were taken using two telescopes of the 
Las Cumbres Global Telescope Network. However, the coverage of the follow-up 
observation was too short (several hours) to give constraints on modeling the 
light curve and thus we do not use the data in the analysis. After the anomaly, 
the light curve gradually declined. Besides the central anomaly, the event is 
different from typical lensing events in the sense that the duration of the event 
is very long. The lensing-induced brightening of the source star started from the 
beginning of the 2016 bulge season and continued even after the end of the season.

Photometry of the data was conducted using codes based on the difference imaging 
method \citep{Tomaney1996, Alard1998} and customized by the individual 
groups: \citet{Udalski2003}, \citet{Bond2001}, and \citet{Albrow2009} for the OGLE, 
MOA, and KMTNet surveys, respectively. For the analysis of data taken by different 
telescopes and processed by different photometry codes, we normalize the error bars 
of the individual data sets following the usual procedure described in \citet{Yee2012}, i.e.
\begin{equation}
\sigma=k(\sigma_0^2+\sigma_{\rm min}^2)^{1/2}.
\end{equation}
Here $\sigma_0$ represents the error bar estimated from the photometry pipeline, 
$\sigma_{\rm min}$ is a term used to adjust error bars so that error bars become consistent 
with their scatter, and $k$ is a factor used to make the $\chi^2$ per degree of 
freedom unity. The $\chi^2$ value is calculated based on the best-fit solution of 
the lensing parameters obtained from modeling. In Table~\ref{table:one}, we list 
the error-bar adjustment factors for the individual data sets.

\begin{deluxetable}{lcc}
\tablecaption{Error-bar adjustment factors \label{table:one}}
\tablewidth{0pt}
\tablehead{
\multicolumn{1}{c}{Data set} &
\multicolumn{1}{c}{$k$} &
\multicolumn{1}{c}{$\sigma_{\rm min}$} 
}
\startdata
OGLE  &  1.304  &  0.005    \\
MOA   &  1.241  &  0.010    \\ 
KMTC  &  1.044  &  0.010    \\ 
KMTS  &  1.109  &  0.020    \\
KMTA  &  1.572  &  0.020   
\enddata                                              
\end{deluxetable}

\section{Modeling Light Curve}

When a lens is composed of two masses, the lens system induces a network of 
caustics at which the lensing magnification of a point source becomes infinity. 
Therefore, the light curve of a binary-lens event with a source trajectory that 
crosses or approaches close to the caustic results in deviations from the smooth 
light curve of a point-mass event. Caustics of a binary lens form a single or 
multiple sets of closed curves.

A central anomaly in the lensing light curve of a high-magnification event 
occurs in two cases of binary lenses. The first case is a binary composed of 
roughly equal masses with a projected separation either substantially smaller 
(close binary) or larger (wide binary) than the angular Einstein radius $\thetae$. 
For a close binary, there exist three sets of caustics where one is located 
around the barycenter of the binary lens and the other two are located away 
from the center of mass. Then, a central anomaly of a high-magnification 
event occurs when a source passes close to the central caustic. For a wide 
binary, on the other hand, there exist two sets of caustics located close 
to the individual lens components. In this case, a central anomaly occurs 
when a source approaches either of the two caustics. The other binary-lens case 
producing central anomalies is a binary composed of two masses with extreme 
mass ratios, e.g. star-planet systems. In the case, the low-mass lens component 
induces a small caustic near the high-mass lens component and the central 
perturbation occurs when the source passes close to the caustic around the 
high-mass component.

Knowing the possible causes of central anomalies, we search for a solution of 
the binary-lens parameters that best describe the observed lensing light curve. 
Under the assumption that the relative lens-source motion is rectilinear, the 
light curve of a binary-lens event is described by 7 principle parameters. 
These parameters include the time of the closest source approach to a 
reference position of the lens, $t_0$, the lens-source separation at that 
moment, $u_0$ (impact parameter), and the timescale for a lens to cross the 
Einstein radius, $t_{\rm E}$ (Einstein timescale), the projected lens-source 
separation normalized to $\thetae$, $s$, the mass ratio between the lens 
components, $q$, the angle between the source trajectory and the binary-lens axis, 
$\alpha$ (source trajectory angle), and the ratio of the angular source radius 
$\theta_*$ to the angular Einstein radius, $\rho=\theta_*/\thetae$ (normalized source 
radius). We note that the normalized source radius is needed to account for 
finite-source effects that occur when the source approaches a caustic. For 
the reference position on the lens plane, we use the barycenter for a 
close binary lens and the photocenter for a wide binary lens. The photocenter 
is located on the binary axis with an offset $q/[s(1+q)]$ from each lens 
component \citep{An2002}.

Since the central anomaly is likely to be produced by the source approach 
close to a caustic, we consider light curve variation caused by finite-source 
effects. We compute finite-source magnifications by applying both numerical 
and semi-analytic approaches. In the numerical approach, we use the 
ray-shooting method. In this method, a large number of rays are uniformly 
shot from the image plane, bent according to the lens equation, and 
gathered on the source plane. With the map of rays on the source plane, 
the finite-source magnification for a given source position is computed 
as the number density ratio of rays on the source surface to the density 
on the image plane. The ray-shooting method is computer intensive. To 
accelerate computation, we compute magnifications in the vicinity of the 
regions around caustics using the semi-analytic hexadecapole approximations 
\citep{Pejcha2009, Gould2008}. In computing finite magnifications, we 
take account the limb-darkening effects of the source star by using the model of 
the surface brightness profile 
$S_\lambda \propto 1-\Gamma_\lambda (1-3 \cos \phi/2)$, 
where $\phi$ is the angle between the light of sight toward the center of 
mass and the normal to the source surface. The limb-darkening coefficients 
$\Gamma_\lambda$ are adopted from \citet{Claret2000} considering the stellar 
type of the source star. See Section 4 for the procedure to decide the 
source type. The adopted coefficients are $\Gamma_I=0.45$ and 
$\Gamma_V=0.62$. For the non-standard MOA filter, we adopt a mean 
value between $R$ and $I$ band, i.e.\ $\Gamma_{\rm MOA}=0.53$.

Searching for solutions of the lensing parameters is conducted in multiple steps. 
In the first step, we conduct a dense grid search over the $(s,q,\alpha)$ 
parameter space while other parameters are search for using a downhill 
approach. We choose $s$, $q$, and $\alpha$ as grid parameters because lensing 
magnifications vary sensitively to small changes of these parameters, while 
the magnification variation to the changes of the other parameters is
smooth. We use the Markov Chain Monte Carlo (MCMC) method for the downhill 
approach. This preliminary search provides local minima in the parameter 
space. In the second step, we refine the individual local solutions first 
by narrowing down the grid space and then allowing all parameters to vary. 
In the final step, we find a global solution by comparing $\chi^2$ value of 
the local solutions. Considering that the central perturbation in the 
light curve of OGLE-2016-BLG-1469 can be produced by either a close/wide 
binary or a planetary companion, we set the ranges of $s$ ($-1.0< \log s <1.0$) 
and $q$ ($-5.0 < \log q < 1.0$) wide enough to investigate both possibilities.

\begin{figure}
\includegraphics[width=\columnwidth]{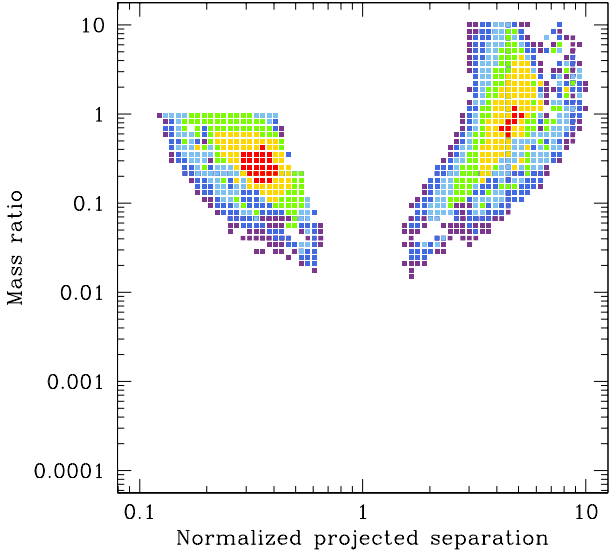}
\caption{$\Delta\chi^2$ map 
of the MCMC chain in the parameter space of the projected separation $s$ 
and the mass ratio $q$ between the binary components. Color coding represents 
the regions with $\Delta\chi^2 < n$ (red), $4n$ (yellow), $9n$ (green), 
$16n$ (cyan), $25n$ (blue), and $36n$ (purple), where $n=4$.}
\label{fig:two}
\end{figure}

In the initial modeling based on the principal parameters (standard model), 
we find no solution that can adequately describe the observed light curve, 
especially the asymmetric feature of the light curve.  See the residual of 
the standard model presented in the bottom panel of Figure~\ref{fig:one}.
Considering that the 
duration of the event comprises a significant portion of the Earth's 
orbital period, the deviation of the lens-source relative motion from 
rectilinear could be significant to cause long-term deviations in the 
lensing light curve, i.e.\ parallax effect \citep{Gould1992}. We, therefore, 
conduct an additional search for solutions by taking the parallax 
effect into account. Including the parallax effect in modeling requires 
to add two more parameters $\pien$ and $\piee$, which represent the components 
of the microlens parallax vector $\pivec_{\rm E}$ projected onto the sky along 
the north and east equatorial coordinates, respectively. The microlens parallax 
vector is related to the relative lens-source parallax 
$\pi_{\rm rel}={\rm au}(D_{\rm L}^{-1}-D_{\rm S}^{-1})$ and the angular 
Einstein radius by
\begin{equation}
\pivec_{\rm E}={\pi_{\rm rel} \over \thetae} {\muvec \over \mu},
\end{equation}
where $\muvec$ represents the relative lens-source proper motion. 
From the modeling including the parallax effect, we find models 
that adequately describe the observed light curve.

Figure~\ref{fig:two} shows the $\Delta\chi^2$ map of the MCMC chain 
in the $s$ -- $q$ parameter 
space obtained from the grid search including parallax effects. We 
identify two locals in the close ($s<1$) and wide ($s>1$) binary regimes, 
which are caused by the well-known close/wide degeneracy 
\citep{Dominik1999, Bozza2000, An2005}. The map also shows that the 
central anomaly in the observed light curve was produced by a binary with 
roughly equal mass components rather than a planetary system.

It is known that the orbital motion of a binary lens can also induce 
long-term deviations in lensing light curves 
\citep{Dominik1998, Albrow2000, Park2013}. We, therefore, 
conduct an additional modeling of the light curve taking account of 
the lens-orbital effects. To the first-order approximation, the 
lens-orbital effect is described by two parameters $ds/dt$ and $d\alpha/dt$, 
which represent the rates of change of the binary separation and the 
source-trajectory angle, respectively.

\begin{deluxetable}{llcc}
\tablecaption{Comparison of Models \label{table:two}}
\tablewidth{0pt}
\tablehead{
\multicolumn{2}{c}{Model} &
\multicolumn{2}{c}{$\chi^2$} \\
\multicolumn{2}{c}{} &
\multicolumn{1}{c}{Close} &
\multicolumn{1}{c}{Wide} 
}
\startdata
Standard       &            &   1849.0   &   1778.8   \\
Parallax       & ($u_0>0$)  &   1384.5   &   1396.1   \\ 
               & ($u_0<0$)  &   1385.2   &   1395.7   \\ 
Orbit+Parallax & ($u_0>0$)  &   1380.3   &   1393.4   \\
               & ($u_0<0$)  &   1381.0   &   1393.0        
\enddata                                              
\end{deluxetable}

\begin{figure}
\includegraphics[width=\columnwidth]{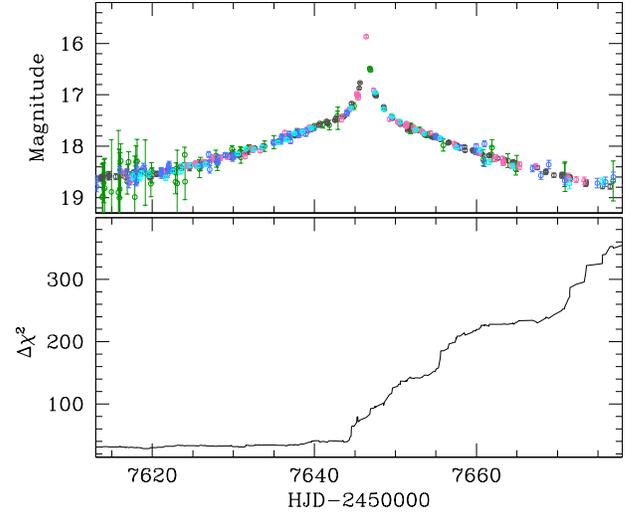}
\caption{
Cumulative distribution of $\Delta\chi^2$ between the best fit-model considering 
higher-order effects and the standard model. The light curve in the upper panel 
is presented to show the region of $\chi^2$ difference.
}
\label{fig:three}
\end{figure}

In Table~\ref{table:two}, we present $\chi^2$ values of the models that we tested. 
Here the ``standard'' model represents the solution based on the 7 principle 
binary-lens parameters. The ``parallax'' model denotes the solution found 
by considering the parallax effect, and the ``orbit+parallax'' model 
represents the solution obtained by considering both the orbital and parallax 
effects. The notations ``close'' and ``wide'' denote the pair of solutions 
with $s<1$ and $s>1$, respectively. 
The models denoted by ``$u_0>0$'' and ``$u_0<0$'' represent the pair of solutions 
resulting from the ``ecliptic degeneracy'', which is caused by the mirror symmetry 
of the source trajectory with respect to the binary axis \citep{Skowron2011}.

From the comparison of the models, it is found that the parallax effect significantly 
improves the fit, i.e.\ $\Delta\chi^2=464.5$ and 382.7 for the close and wide-binary 
cases, respectively. On the other hand, the improvement by additionally considering the 
lens-orbital effect, $\Delta\chi^2<5$, is meager. We also find that the close-binary 
solution is moderately ($\Delta\chi^2 > 10$) preferred over the wide-binary solution. 
However, the degeneracy between the $u_0>0$ and $u_0<0$ solutions is very severe 
($\Delta\chi^2 < 1$). In Figure~1, we present the model light curve of the best-fit 
solution, i.e.\ close ``orbit+parallax'' model with $u_0>0$, plotted over the data 
points. For comparison, we present the residuals from the best-fit solution and the 
standard solution in the lower two panels. In Figure~\ref{fig:three}, we also present 
the cumulative distribution of $\Delta\chi^2$ between the best-fit solution and the 
standard model to show the region of $\chi^2$ improvement.  One finds that $\chi^2$ 
improvement occurs during the central anomaly and throughout the region after the peak.

\begin{deluxetable}{lrr}
\tablecaption{Best-fit lensing parameters \label{table:three}}
\tablewidth{0pt}
\tablehead{
\multicolumn{1}{c}{Parameters} &
\multicolumn{1}{c}{$u_0>0$}    &
\multicolumn{1}{c}{$u_0<0$}  
}
\startdata
$t_0$ (HJD')                     &  7646.598 $\pm$ 0.035   &  7646.621  $\pm$ 0.034      \\  
$u_0$                            &     0.052 $\pm$ 0.001   &    -0.051  $\pm$ 0.001      \\    
$\te$ (days)                     &     99.74 $\pm$ 0.76    &     99.35  $\pm$ 0.84       \\     
$s$                              &     0.354 $\pm$ 0.010   &     0.345  $\pm$ 0.011      \\  
$q$                              &     0.269 $\pm$ 0.020   &     0.289  $\pm$ 0.029      \\    
$\alpha$ (rad)                   &    -0.073 $\pm$ 0.004   &     0.063  $\pm$ 0.009      \\ 
$\rho$ ($10^{-3}$)               &     4.48  $\pm$ 0.41    &     4.33   $\pm$ 0.49       \\  
$\pi_{{E},N}$                    &     0.226 $\pm$ 0.045   &    -0.179  $\pm$ 0.051      \\   
$\pi_{{E},E}$                    &     0.360 $\pm$ 0.013   &     0.367  $\pm$ 0.012      \\     
$ds/dt$ (${\rm yr}^{-1}$)        &    -0.390 $\pm$ 0.048   &    -0.273  $\pm$ 0.066      \\     
$d\alpha/dt$ (${\rm yr}^{-1}$)   &    -0.052 $\pm$ 0.061   &    -0.099  $\pm$ 0.070      \\    
$(F_s/F_b)_{{\rm OGLE},I}$       &     0.120/0.143         &     0.119/0.143             
\enddata                                              
\tablecomments{${\rm HJD}'={\rm HJD}-2450000$.}
\end{deluxetable}

In Table~\ref{table:three}, we present the lensing parameters of the best-fit 
solutions.  Due to the severity of the degeneracy, we present both the 
$u_0>0$ and $u_0<0$ solutions. Also presented is the fluxes from the source, 
$F_s$, and blend, $F_b$, that are measured in the $I$-band OGLE data.
The uncertainty of each parameter is estimated based on the scatter of 
points in the MCMC chain.

\begin{figure}
\includegraphics[width=\columnwidth]{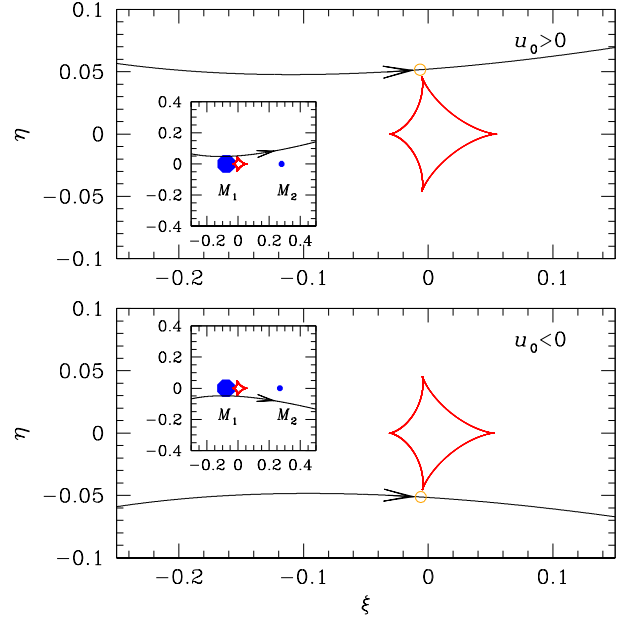}
\caption{Geometry of the lens system. The upper and lower panels correspond 
to the $u_0>0$ and $u_0<0$ solutions, respectively. The coordinates $(\xi,\eta)$ 
are centered at the barycenter of the lens system and lengths are normalized to 
the angular Einstein radius. The cuspy close figure represents the caustic and 
the curve with an arrow is the source trajectory. The small orange circle at 
the tip of the arrow on the source trajectory represents the source size 
relative to the caustic. The inset in each panel is inserted to show the 
caustic position with respect to the binary-lens components (marked by blue 
dots).}
\label{fig:four}
\end{figure}

\begin{figure}
\includegraphics[width=\columnwidth]{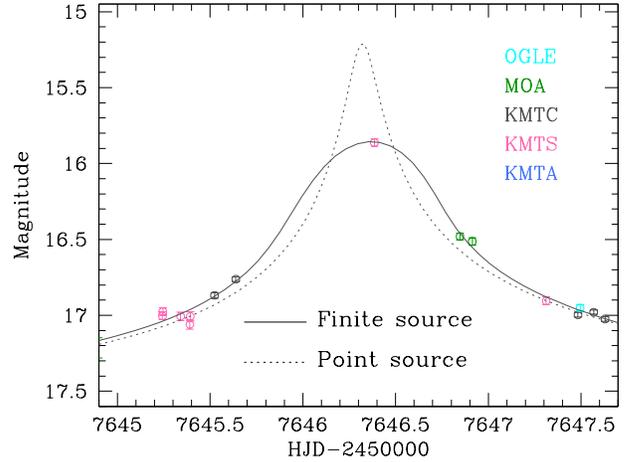}
\caption{Model light curves around the central anomaly region resulting from a 
finite and a point source.  }
\label{fig:five}
\end{figure}

In Figure~\ref{fig:four}, we present the lens system geometry which shows 
the source trajectory (the curve with an arrow) with respect to the lens 
components (the blue dots) and caustics (the cuspy close curve). 
We present two cases corresponding to the $u_0>0$ (upper panel) and $u_0<0$ 
(lower panel) solutions.
We find 
that the event was produced by a binary composed of two masses with a 
projected separation $s\sim 0.35$ and a mass ratio $q\sim 0.28$. The binary 
lens induced a small 4-cusp caustic around the barycenter of the lens and 
the source star moved almost in parallel with the binary-lens axis, i.e.\ 
$\alpha\sim 0$. The central anomaly was produced when the source passed the 
tip of the off-binary-axis cusp. To be mentioned is that although the source 
star did not cross the caustic, finite-source effects are clearly detected. 
This is possible due to the steep magnification gradient in the region 
extending from the strong cusp of the caustic. In Figure~\ref{fig:five}, we 
present the two model light curves resulting from a finite (solid curve) and 
a point source (dotted curve).  We note that the point-source light curve is 
based on the same lensing parameters as those of the finite-source model 
except $\rho$.

\begin{figure}
\includegraphics[width=\columnwidth]{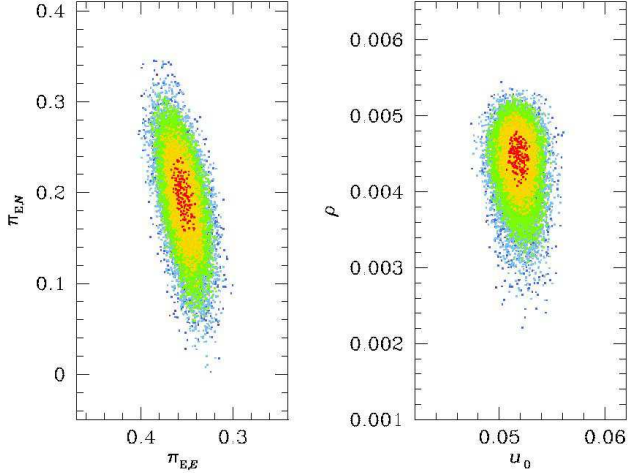}
\caption{$\Delta\chi^2$ distributions of MCMC chains in the $\piee$ --$\pien$ 
(left panel) and $u_0$ -- $\rho$ (right panel) parameter space. Dots in different 
color represent chains with $\Delta\chi^2 < 1$ (red), 4 (yellow), 9 (green), 
16 (cyan), and 25 (blue).}
\label{fig:six}
\end{figure}

\section{Physical Lens Parameters}

As shown in Figure~\ref{fig:six}, where we present the $\Delta\chi^2$ 
distributions of MCMC chains in the $\piee$ -- $\pien$ (left panel) and 
$u_0$ -- $\rho$ (right panel) parameter space, both microlens-parallax and 
finite-source effects are clearly detected. In this case, one can measure both 
$\pie=(\pien^2+\piee^2)^{1/2}$ and $\thetae$, which are the two ingredients 
needed for the  unique determinations of the mass $M$ and the distance to 
the lens $D_{\rm L}$, i.e. 
\begin{equation}
M={\thetae \over \kappa\pie};\qquad
D_{\rm L}={{\rm au} \over \pie\thetae +\pi_{\rm S}}.
\end{equation}
Here $\kappa=4G/(c^2 {\rm au})$, $\pi_{\rm S}={\rm au}/D_{\rm S}$ denotes the 
parallax of the source, and $D_{\rm S}$ is the distance to the source.

\begin{figure}
\includegraphics[width=\columnwidth]{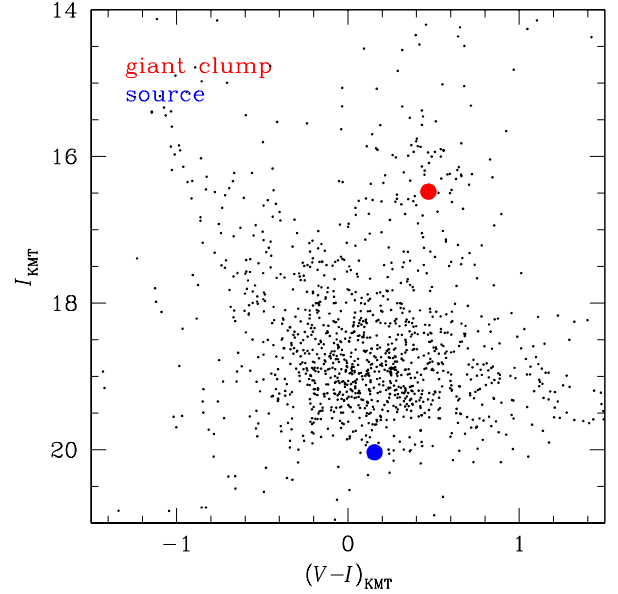}
\caption{Position of the source star with respect to the centroid of giant 
clump in the instrumental color-magnitude diagram.}
\label{fig:seven}
\end{figure}

The angular Einstein radius is determined from the combination of the measured 
finite-source parameter $\rho$ and the angular radius of the source star, 
$\theta_*$, i.e.\ $\thetae= \rho/\theta_*$. We determine the angular source size 
based on the source type deduced from the color and brightness. In order 
to estimate the de-reddened color $(V-I)_0$ and brightness $I_0$ of the source 
star, we use the usual method of \citet{Yoo2004}, where $(V-I)_0$ and $I_0$ are 
estimated based on the offsets in color and magnitude from those of the centroid 
of giant clump for which the de-reddened color and  brightness 
$(V-I,I)_{{\rm GC},0}=(1.06,14.25)$ \citep{Bensby2011, Nataf2013} are known. 
Figure~\ref{fig:seven} shows the location of the source star with respect to 
the giant clump centroid in {\it instrumental} the color-magnitude diagram of 
stars in the neighboring region around the source.  We note that the color-magnitude 
diagram is uncalibrated and thus one cannot determine the reddening $E(V-I)=(V-I)-(V-I)_0$ 
or the extinction $A_I=I-I_0$.  However, one can measure the de-reddened color and 
magnitude of the source star based on the offsets $\Delta(V-I,I)=(-0.3,3.5)$ with 
respect to the centroid of the giant clump. The estimated de-reddened color and 
magnitude of the source star are $(V-I,I)_0=(0.8,18.5)$ and find that the source is 
a mid G-type main-sequence star.  Once $(V-I)_0$ color is estimated, we convert into 
$(V-K)_0= 1.64\pm 0.05$ using the $VI/VK$ relation \citep{Bessell1988} and then 
estimate the source angular radius from the $VK/\theta_*$ relation \citep{Kervella2004}.  
The estimated angular source radius is $\theta_*=0.94 \pm 0.07\ \mu{\rm as}$.  The 
angular Einstein radius estimated from $\theta_*$ and $\rho$ is
\begin{equation}
\thetae= 
\cases{
0.21 \pm 0.03\ {\rm mas} & for $u_0>0$, \cr     
0.22 \pm 0.03\ {\rm mas} & for $u_0<0$. \cr
}
\end{equation}
With the measured timescale, then, the relative lens-source proper 
motion is estimated by $\mu=\thetae/t_{\rm E}$. We find that 
\begin{equation}
\mu=
\cases{
0.87 \pm 0.10\ {\rm mas}\ {\rm yr}^{-1}  &  for $u_0>0$, \cr
0.81 \pm 0.11\ {\rm mas}\ {\rm yr}^{-1}  &  for $u_0<0$. \cr
}
\end{equation}
We note the measured relative proper motion is substantially smaller 
than $\sim 5\ {\rm mas}\ {\rm yr}^{-1}$ of typical Galactic lensing events. 
We discuss the probability of low proper motions in Section 5.

\begin{deluxetable}{lcc}
\tablecaption{Physical lens parameters \label{table:four}}
\tablewidth{0pt}
\tablehead{
\multicolumn{1}{c}{Parameter} &
\multicolumn{2}{c}{Model} \\
\multicolumn{1}{c}{} &
\multicolumn{1}{c}{Close} &
\multicolumn{1}{c}{Wide} 
}
\startdata
$M_1$ ($M_\odot$)            &  $0.048 \pm 0.007$   &   $0.051 \pm 0.008$  \\
$M_2$ ($M_\odot$)            &  $0.013 \pm 0.002$   &   $0.015 \pm 0.002$  \\
$D_{\rm L}$ (kpc)            &  $4.47 \pm 0.51  $   &   $4.48 \pm 0.51  $  \\
$a_\perp$ (au)               &  $0.33 \pm0.04   $   &   $0.34 \pm 0.04  $  \\
$({\rm KE}/{\rm PE})_\perp$  &  0.01                &   0.01         
\enddata                                              
\end{deluxetable}

In Table~\ref{table:four}, we present the estimated physical lens parameters. 
Here $M_1$ and $M_2$ denote the masses of the lens components, 
$a_\perp=s D_{\rm L}\thetae$ is the physical size of the projected separation 
between the lens components, and $({\rm KE}/{\rm PE})_\perp$ denotes the 
transverse kinetic-to-potential energy ratio that is defined by \citep{Dong2009}
\begin{equation}
\left( {{\rm KE}\over{\rm PE}}\right)_\perp
={(a_\perp/{\rm au})^3\over 8 \pi^2 (M/M_\odot)}
\left[  
\left( {1\over s} {ds/dt  \over {\rm yr}^{-1}}\right)^2+
\left( {d\alpha/dt  \over {\rm yr}^{-1}}\right)^2
\right].
\end{equation}
This ratio is less than the three-dimensional kinetic-to-potential energy 
ratio, KE/PE, which is less than unity for a bound system, i.e.\ 
$({\rm KE}/{\rm PE})_\perp < {\rm KE}/{\rm PE} < 1$. We find that both lens 
components have masses below the hydrogen-burning limit of $\sim 0.08\ M_\odot$, 
indicating that the lens is a brown-dwarf binary. This is the third microlensing 
brown-dwarf binary followed by the first discoveries of OGLE-2009-BLG-151L and 
OGLE-2011-BLG-0420L by \citet{Choi2013}.  The brown-dwarf binary is located at 
a distance $D_{\rm L}\sim 4.5$ kpc from the Earth.  The projected separation 
between the binary components is $a_\perp=0.33$ au and 0.34 au for the $u_0>0$ 
and $u_0<0$ solutions, respectively.

\section{Discussion}

As mentioned in the previous section, the measured relative lens-source proper 
motion $\mu\sim 0.8$ -- 0.9 mas ${\rm yr}^{-1}$ is substantially smaller than 
$\sim 5\ {\rm mas}\ {\rm yr}^{-1}$ of typical Galactic lensing events. Tracking 
down the cause of the low $\mu$ value is important because $\mu$ is estimated 
from $\thetae$ and thus wrong $\thetae$ determination can lead to erroneous 
determinations of the other physical lens parameters.

In order to trace the origin of the estimated low proper motion, we conduct a Monte 
Carlo simulation of Galactic microlensing events based on the models of physical and 
dynamical distributions of the Galaxy combined with the mass function of objects. 
We adopt the Galactic model of \citet{Han1995} for the physical and dynamical 
distributions of matter. 
In this model, the disk matter distribution is modeled by a double-exponential
disk and the velocity distribution is assumed to be Gaussian about the rotation velocity.
The bulge is modeled by a triaxial bulge and 
matter in the bulge moves following 
a triaxial Gaussian distribution, where the velocity components 
along the major axes are deduced from
the flattening of the bulge via the tensor virial theorem.
For the mass function of lens objects, we adopt the 
\citet{Gould2000} model, which is constructed based on a stellar luminosity function 
plus stellar remnants.
Based on the models, we produce a large number ($10^5$) 
of mock events and compute timescales and proper motions of events.

\begin{figure}
\includegraphics[width=\columnwidth]{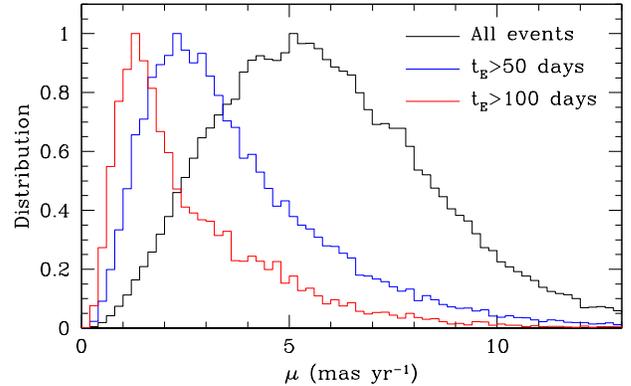}
\caption{Distributions of relative lens-source proper motion for three different 
populations of Galactic lensing events: all events (black curve), events with 
$t_{\rm E}>50$ days (blue curve), and those with $t_{\rm E}>100$ days.}
\label{fig:eight}
\end{figure}

In Figure~\ref{fig:eight}, we present the distributions of relative lens-source 
proper motions obtained from the simulation. To see the variation of the 
proper-motion distribution $f(\mu)$ depending on the event timescale, we produce 
distributions for three different populations of events: (1) all events, (2) 
events with timescales $t_{\rm E}>50$ days, and (3) those with $t_{\rm E}>100$ 
days. From the presented distributions, it is found that the proper-motion 
distributions show a trend where  the mode value of the $\mu$ distribution 
becomes smaller as the event timescale increases. As expected, the mode value 
of all Galactic lensing events is $\mu\sim 5\ {\rm mas}\ {\rm yr}^{-1}$. 
However, the relative proper motion becomes smaller with the increase of the 
event time scale.  It is found that the mode value of the distributions are 
$\mu \sim 2.3\ {\rm mas}\ {\rm yr}^{-1}$ and $\sim 1.2\ {\rm mas}\ {\rm yr}^{-1}$ 
for events with $t_{\rm E}>50$ days and $>100$ days, respectively.  Considering 
that the timescale of OGLE-2016-BLG-1469 is $\sim 100$ days, therefore, the 
measured proper motion of $\mu\sim 0.8$ -- 0.9 mas ${\rm yr}^{-1}$ is not an 
abnormally small value but a typical value for long time-scale events.

Being able to detect brown-dwarf binaries that are difficult to be detected by 
other methods, microlensing is a complementary method to other methods.  The 
majority of nearby brown-dwarf binaries discovered by direct imaging have 
separations in the range 1 to 10 AU. See the histogram of projected separations 
presented in Figure 2 of \citet{Aller2012}.  Microlensing sensitivity is maximum 
for binaries separated by the Einstein radius, which is related to the mass and 
distance of the binary by
\begin{equation}
r_{\rm E}\sim 0.9\ {\rm au} \left( {M_{\rm tot}\over 0.05\ M_\odot}  \right)^{1/2}  
\left( {D_{\rm S} \over 8\ {\rm kpc}}\right)^{1/2}
\left[ {x(1-x)\over 0.25} \right]^{1/2},
\label{eq7}
\end{equation}
where $x=D_{\rm L}/D_{\rm S} < 1.0$.  Therefore, microlensing is sensitive to 
close brown-dwarf binaries with separations $a_\perp \lesssim 1$ au, for which 
the sensitivities of other methods  are low.  Actually, the projected separations 
of the three microlensing brown-dwarf binaries are $a_\perp\sim 0.3$ au, 0.2 au, 
and 0.3 au for OGLE-2009-BLG-151L, OGLE-2011-BLG-0420L, and this system, respectively. 
The mass ratio histogram of brown-dwarf binaries discovered by direct imaging
\citep{Aller2012} exhibits a clear tendency of equal mass systems. The lack of 
low-mass ratio systems is likely to be due to the difficulty in detecting second 
components.  By contrast, the dependency of the microlensing sensitivity to the 
mass ratio is weak and the efficiency extends down to planetary companions with 
$q < 0.1$, e.g. OGLE-2012-BLG-0358Lb \citep{Han2013}.
There exist multiple theories for the formation of binary brown dwarfs, 
e.g.\ \citet{Reipurth2001}, \citet{Stamatellos2009}, and \citet{Bate2009}.
These different formation mechnisms would result in different binary properties 
such as binary frequency, mass ratios and separations.  Therefore, a sample 
comprising brown-dwarf binaries with physical parameters spanning wide ranges is 
important to better understand the formation mechanism of binary brown dwarfs.

\section{Conclusion}
We reported the microlensing discovery of a binary that was composed of two 
brown dwarfs.  The brown-dwarf binary was found from the analysis of the 
microlensing event OGLE-2016-BLG-1469.  The light curve of the event exhibited 
a short-term central anomaly, which turned out to be produced by a binary 
companion with a mass roughly equal to the primary.  Although the source star 
did not cross the caustic induced by the binary companion, finite-source 
effects were clearly detected, enabling us to measure the angular Einstein 
radius.  In addition, we measures the microlens parallax from the the asymmetric 
light curve.  By measuring both the angular Einstein radius and the microlens 
parallax, we could uniquely determine the masses and identified the substellar 
nature of the lens components.  The lens was the third microlensing brown-dwarf 
binary with measured mass, demonstrating the usefulness of the microlensing 
method in detecting field brown-dwarf binaries.

\begin{acknowledgments}
Work by C.~Han was supported by the Creative Research Initiative Program (2009-0081561) of 
National Research Foundation of Korea.  
The OGLE project has received funding from the National Science Centre, Poland, grant 
MAESTRO 2014/14/A/ST9/00121 to A.~Udalski.  OGLE Team thanks Profs.\ M.~Kubiak, G.~Pietrzy{\'n}ski, 
and {\L}.~Wyrzykowski for their contribution to 
the collection of the OGLE photometric data over the past years.
The MOA project is supported by JSPS KAKENHI Grant Number JP24253004, JP26247023, 
JP16H06287, JP23340064 and JP15H00781.
Work by A.~Gould was supported by JPL grant 1500811.
A.~Gould and W.~Zhu acknowledges the support from NSF grant AST-1516842. 
Work by J.~C.~Yee. was performed under contract with
the California Institute of Technology (Caltech)/Jet Propulsion
Laboratory (JPL) funded by NASA through the Sagan
Fellowship Program executed by the NASA Exoplanet Science
Institute.
This research has made use of the KMTNet system operated
by the Korea Astronomy and Space Science Institute (KASI)
and the data were obtained at three host sites of CTIO in
Chile, SAAO in South Africa, and SSO in Australia. 
We acknowledge the high-speed internet service (KREONET)
provided by Korea Institute of Science and Technology Information (KISTI).

\end{acknowledgments}

\end{document}